\documentclass[twocolumn,aps,prl,showpacs,amsmath,amssymb,psfrac]{revtex4}

\usepackage{amsmath}

\usepackage{epsfig}
\usepackage{graphicx}
\usepackage{dcolumn}
\usepackage{bm}
\begin{document}

\title{Interactions and magnetism in graphene boundary states}

\author{B. Wunsch$^{1,2}$, T. Stauber$^{2,3}$, F. Sols$^{1}$ and F. Guinea$^{2}$}
\affiliation{$^1$ Departamento de F\'isica de Materiales, Universidad Complutense de Madrid, E-28040
  Madrid, Spain.\\$^2$ Instituto de Ciencia de Materiales de Madrid, CSIC,
  E-28049 Madrid, Spain.}
\affiliation{$^3$Departamento de
F\'{\i}sica, Universidade do Minho, P-4710-057, Braga, Portugal}

\date{\today}
\begin{abstract}
  We analyze interaction effects on boundary states of single layer graphene.
  Near a half filled band, both short and long-ranged interactions lead to a
  fully spin polarized configuration. In addition, the band of boundary states acquires a
  finite dispersion as function of the momentum parallel to the edge,
  induced by the interactions. Away from half filling the wavefunction
  develops charge correlations similar to those in a Wigner crystal, and the
  spin strongly alternates with the occupation of the boundary states. For
  certain fillings the ground state has a finite linear momentum,
  leading to the formation of persistent currents.
\end{abstract}
%
\maketitle

{\em Introduction.} The relativistic low energy properties of graphene give rise to many
anomalies with respect to semiconductor physics and are thus very interesting
from a fundamental point of view~\cite{rmp,GN07}.
Furthermore, carbon-based nanoelectronics have attracted much interest as they
might complement silicon-based devices~\cite{Avouris07}. An important
question concerns the magnetic properties of graphene.  While ideal graphene
sheets are far away from the ferromagnetic phase transition~\cite{PGN05}, the
occurrence of midgap states, that lead to a peak in the density at the Dirac
points, can lead to magnetism.

The most studied example of midgap states are boundary
states~\cite{FWNK96,Nakada96,Wak99}.  Depending on the kind of edge there is a
mismatch between the number of $A$ and $B$ sites and Lieb theorem (strictly
applicable only to Hubbard interaction) guarantees a magnetic ground state at
half filling with a spin given by $S=(N_A-N_B)/2$~\cite{Lieb}.  A prominent
example are zigzag-edges where the outermost atom corresponds always to the
same site.  However, boundary
states are present not only for zigzag edges but for any boundary except for
pure armchair edges~\cite{Nakada96,Akhmerov07}.  The originally flat energy band of
boundary states is strongly affected by electron-electron interactions and first principle calculations predict magnetic
boundaries~\cite{Son06,Rossier07,Yazyev07,Sasaki08}.

Previous approaches mostly consider the short-ranged Hubbard interaction. 
However, in graphene the screening close to
half filling is known to be poor~\cite{DiVincenzo84} and a comparison with
results considering the more realistic case of long-ranged Coulomb interactions is therefore desirable.
For extended systems and finite doping, this is typically done within the
random phase approximation~\cite{Wunsch06,Hwang07}.  The effect of
interaction on the flat band of midgap states however is better described by
non-perturbative treatments~\cite{Wunsch08}.

In this work we present a simple effective model which allows us to assess the
long-ranged Coulomb interaction as well as Hubbard interaction. The basic
assumption is the existence of a filled valence and an empty conduction band
whose properties do not depend on the filling of the boundary states. The model
can be justified by a)~extended states are separated from the degenerate
boundary states by an energy gap for any confined system, b)~minimization of the
classical electrostatic energy favors the occupancy of the boundary states, c) at
low energies the density of states is strongly dominated by the boundary states.
We note that Hartree and
exchange interactions between valence band and boundary states are taken into account.
Finally, the interaction between electrons occupying boundary states is
treated exactly via exact diagonalization of the corresponding Hamiltonian.

Our results are the following: (i)~Both interactions give similar results
close to half filling, however, there are deviations away from half filling,
(ii)~Our effective model conserves all symmetries of the full Hamiltonian and
can be expressed in terms of low energy states only, (iii) The initially flat
 band of boundary states acquires a finite dispersion due to interaction effects, (iv)~There
is a finite regime around half-filling, where the edge is maximally polarized,
i.e., the edge is magnetic, (v) Strong spin features are predicted for low
electron (hole) concentration, which should strongly affect the transport
properties as they may cause spin blockade through graphene islands.  In the
following we first introduce our model and then the main
results are discussed. Finally we present the the conclusions of our work.

\begin{figure}
  \begin{center}
    \includegraphics*[width=0.8\linewidth]{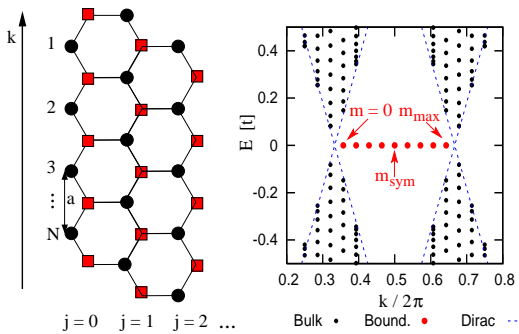} 
    \caption{(Color online) Left: Atomic structure of an unrolled nanotube
      with zigzag edges. 
      $j$ labels the zigzag lines, each containing
      $N$ different $A$ and $B$ sites. $k$
      denotes the wavenumber along the edge. Right: Single-particle spectrum
      for $N=28, N_{B}=9$. Red circles denote boundary states, black circles extended (bulk)
      states and blue dashed line shows the Dirac cone. Boundary states are
      labeled by integer quantum number $m$.  
    }
\label{fig:SParticleNT}
\end{center}
\end{figure}

{\em Edge states.}  While an ideal graphene sheet has a vanishing density of
states at half filling, real samples have localized states at (or around)
zero energy, which are induced close to edges, impurities, defects or wiggles.
As a concrete example of a finite set of localized states, we choose the boundary
states formed at the zigzag edge of a carbon nanotube :~\cite{FWNK96,Nakada96}
\begin{eqnarray}
\phi_{n}&=&\sqrt{1-4\cos^2(k_n/2)}\sum_{j=0}^{\infty} \left[-2\cos(k_n/2)\right]^ja_{jk_n}\;
\label{EdgeStates}
\end{eqnarray} 
Here, $k_n=2 \pi n/N$ denotes the wave number parallel to the edge,
$N$ is the number of $A$-sites at the edge and $a_{jk_n}$ is a
plane wave with momentum $k_n$ living on the $A$-atoms of the $j$-th zigzag
line, see Fig.~\ref{fig:SParticleNT}.  Boundary states exist for all integer
numbers $n$ between $n_{\rm min}=\text{Int}(N/3+1)$ and $n_{\rm max}=\text{Int}((2
N-1)/3)$. From now on the boundary
states are labeled by the integer number $m$, 
which we call angular momentum, $k_m=(n_{\rm min}+m) 2\pi/N$
with $0\leq m\leq m_{\rm max}=n_{\rm max}-n_{\rm min}$ as illustrated in
Fig.~\ref{fig:SParticleNT}. $N_B$ denotes the number of boundary states.

We note that the boundary states of Eq.~(\ref{EdgeStates}) are the zero energy
eigenstates of the original tight binding Hamiltonian describing graphene,
without any linearization around the $K$-points. Thus we also account for
interaction induced scattering between the two $K$-points, which is
particularly important for the boundary states. Since the system is invariant
under exchange of the two $K$-points the energy spectrum is symmetric around
$m_{\rm sym}=m_{\rm max}/2$ as illustrated in Fig.~\ref{fig:SParticleNT}. We
note that boundary states decay fastest for $m=m_{\rm max}/2$.

{\em Interactions.}  We now study the electronic and magnetic properties of
the boundary states if either Hubbard $V_H$ or Coulomb interaction $V_C$ is included:
\begin{eqnarray}
V_{H}&=&U \sum_{i}
\left(n_{i\uparrow}-\frac{1}{2}\right)\left(n_{i\downarrow}-\frac{1}{2}\right)\\
V_C&=&\frac{1}{2}\sum_{i\neq j} \frac{e^2}{r_{ij}} \left(n_i-1\right) \left(n_j-1\right) +V_H
\end{eqnarray}
here $i,j$ denote the lattice sites, $n_i=n_{i\downarrow}+n_{i\uparrow}$ the
occupation of site $i$ with electrons and $\uparrow,\downarrow$ the direction
of the spin. The constant charge density of the positive charge background is
subtracted from the electronic density, which guarantees charge neutrality at
half filling and establishes electron-hole symmetry around half filling. The
Coulomb interaction consists of a long-ranged part (first part with $i \neq
j$) and the Hubbard part where the ratio between both is determined by
$(e^2/a)/U$~\cite{rmin}.

The single particle spectrum of graphene has a sublattice symmetry. To each
states $\psi_v$ with energy $E_v<0$ there exists a state
$\psi_c$ with energy $E_c=-E_v$ and vice versa.  The
states are related by $\psi_c({\bf r}_i)=\epsilon({\bf r}_i)\psi_v({\bf r}_i)$
with $\epsilon({\bf r}_i)=-1$ if ${\bf r_i}$ is a B-site and $\epsilon({\bf
  r}_i)=1$ otherwise. We note that $\psi_v({\bf r})^\star \,\psi_v({\bf
  r'})=\psi_c({\bf r})^\star \psi_c({\bf r'})$.  Therefore we can write the
completeness relation as
\begin{align}
2\sum_v \psi_v({\bf r})^\star \,\psi_v({\bf r'})+\sum_m  \psi_m({\bf
  r})^\star \,\psi_m({\bf r'})=\delta_{\bf r,\bf r'}\label{Eq:ehsym}\,,
\end{align}
where $v$ denotes the (orbital-)index for the valence band and $m$ for the
boundary states.

The interaction between the filled valence band and the
boundary states can be written as $H_{\rm VB}=\sum_mE_mn_m$, where
Eq.~(\ref{Eq:ehsym}) allows to express the potentials $E_m$ in terms of boundary
states only,
\begin{align}
E_m
&=-\sum_n\langle nm|V|nm\rangle+\frac{1}{2}\sum_n\langle
nm|V|mn\rangle\label{OneParticlePotential}\;.
\end{align} 
$V$ is either Hubbard $V_{H}({\bf r_1,r_2})=U \delta_{\bf r_1,r_2}$ or Coulomb
interaction $V_{C}({\bf r_1,r_2})=e^2/|{\bf r_1-r_2}|$.  Since there is no
kinetic energy, the effective Hamiltonian for the boundary states consists
only of the potential term $H_{\rm VB}$ and the mutual interaction between
electrons occupying boundary states.  We note that the sublattice symmetry
allowed us to account for the interaction with the valence band without
calculating the finite energy eigenstates. In fact, even for a finite energy
window around the Dirac points one can express all interactions in terms of
the low energy states within that energy window. The interaction between
electrons occupying boundary states is treated exactly by numerically
diagonalizing the Hamiltonian matrix. Since our effective model conserves the
symmetries of the full Hamiltonian, we can diagonalize the Hamiltonian matrix
in subspaces of given spin, angular momentum and particle number denoted as
$S$, $S_z$, $M$, $N_e$. Since the energy is independent of the spin projection
we set $S_z=S$. Furthermore we note that electron-hole symmetry around half
filling is also conserved.

\begin{figure}
  \begin{center}
    \includegraphics*[width=\linewidth]{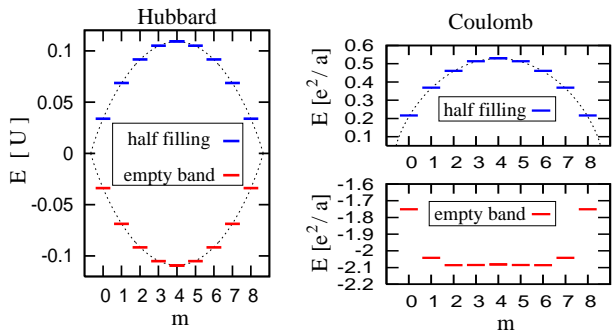} 
    \caption{(Color online) Band dispersion $E_m$ for empty band
      (see Eq.~(\ref{OneParticlePotential})) and half filled band
      $\tilde{E}_m$ (see Eq.~(\ref{OneParticlePotential2})) for $N=28, N_{B}=9$. Dashed lines show
      continuous limit ($N\to \infty$). }
\label{fig:Band}
\end{center}
\end{figure}

Within our model, the empty band of boundary states has zero energy. Since $H_{\rm
  VE}$ is diagonal in the basis of boundary states the eigenenergy of a single
electron occupying a boundary state is given by $E_m$ of
Eq.~(\ref{OneParticlePotential}), which consists of a Hartree term (first
part) and an
exchange term (second part). One can also define a band structure at half filling, where the
ground state is maximally spin polarized. The energy $\tilde{E}_m$ needed to add (or
annihilate) an electron (or hole) to the ground state at half filling by
occupying the boundary state $|m\rangle$ is given by
\begin{align}
\tilde{E}_m&=\frac{1}{2}\sum_n\langle nm|V|mn\rangle\label{OneParticlePotential2}\;.
\end{align}
It only consists of the exchange part. The energy bands are depicted in Fig.~\ref{fig:Band}.

For Hubbard interaction exchange and direct term are the same so that
$E_m=-\tilde{E}_m$ with:
\begin{align*}
E_m=-\frac{U}{2N}\sum_n \frac{\left(1-4\cos^2(k_m/2)\right)\left(1-4\cos^2(k_n/2)\right)}{(1-16\cos^2(k_m/2)\cos^2(k_n/2))} 
\end{align*}
Here $k_m=(n_{\rm min}+m) 2\pi/N$ as explained below Eq.~(\ref{EdgeStates}).
These energies rapidly converge for large $N$ to a band of boundary states as
shown in Fig.~\ref{fig:Band}. The maximum of the band is at
$m=m_{sym}$ ($k=\pi$) and the bandwidth is given by
$U(\sqrt{3}/2\pi-1/6)\approx 0.11 U$. Close to a $K$-point the
dispersion of the boundary states is given by $\pm \hbar v_F k U/3t$,
where $k$ denotes the distance from the $K$-point.
 
For Coulomb interaction, exchange and Hartree term are different. The exchange
interaction is short-ranged and converges fast in the limit of a long edge as
illustrated in Fig.~\ref{fig:Band} for the band at half filling.
However, the Hartree term has a long-ranged contribution that leads to a
$\log(N)$ divergence of the Hartree contribution to the potentials $E_m$.
Already the band structure shown in Fig.~\ref{fig:Band} indicates that 
close to half filling the results obtained by either Hubbard
or Coulomb interaction will resemble each other, while differ results can be
expected close to an empty or filled band of boundary states.

\begin{figure}
  \begin{center}
    \includegraphics*[width=0.8\linewidth]{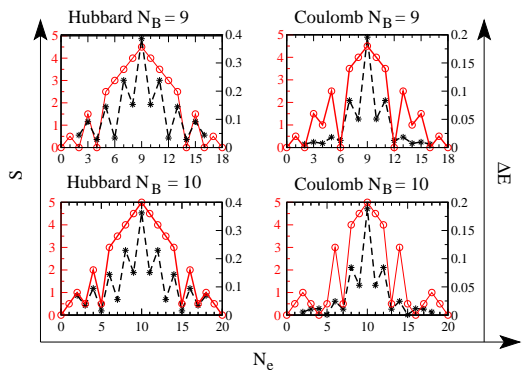} 
    \caption{(Color online) Spin of the ground state (red) and energy gap for
      spin excitations from the ground state (black) as a function of the number
      of boundary electrons.  Upper
      and lower row correspond to $N=28, N_{B}=9$ and $N=31,N_{B}=10$
      respectively. Energies in units of
      $e^2/R$(Coulomb) and $U/N$(Hubbard).}
\label{fig:S0_DE}
\end{center}
\end{figure}

\begin{figure}
  \begin{center}
    \includegraphics*[width=\linewidth]{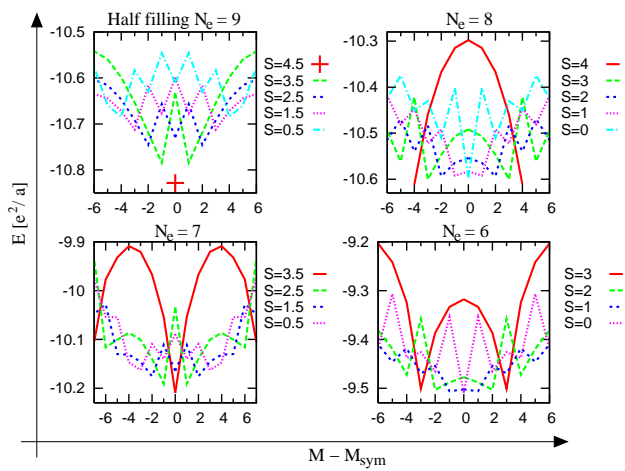} 
    \caption{(Color online) $N_e$-electron ground state energies as
       function of total spin $S$ and angular momentum $M$ ($M_{sym}=N_e
       m_{sym}$). The favored spin depends on the angular momentum. For
       certain occupation numbers (e.g.$N_e=8$) the ground state has a
       finite angular momentum. $N=28, N_{B}=9$.
    }
\label{fig:EMS}
\end{center}
\end{figure}

{\em Results.}  As discussed earlier, the effective model for the boundary
states conserves all symmetries of the full Hamiltonian. We note that it is crucial to include the Hartree and
exchange one-body potentials induced by the interaction with the valence band.

Fig.~\ref{fig:S0_DE} shows the spin of the ground state (red line) and energy
gap for spin excitations (black line) as a function of the number of electrons
occupying boundary states. We find that close to half-filling the system is
maximally polarized, whereas far away from half filling, i.e. close to the
empty or filled band, the addition of a single electron or hole leads to
strong alternations of the spin of the ground state. The energy gap for spin
excitations shows an odd-even effect in the particle number and is generally
bigger close to half filling.

The underlying physics is determined by the competition between the
minimization of the interaction energy on the one hand and of the
single-particle energies on the other hand.  This competition is most obvious
for the Hubbard interaction, where only electrons of opposite spin interact
with each other so that the interaction between boundary states always favors a
spin polarized ground state.  On the other hand the interaction with the
filled valence band leads to a dispersion of the band of boundary states as
illustrated in Fig.~\ref{fig:Band} and a spin unpolarized state allows for a
double occupation of energetically preferred orbitals. 

While close to half
filling the spin polarized state is always favored, we find for a nearly empty
band of boundary states that the reduction in potential energy can exceed the cost
in interaction energy, so that the spin unpolarized state is favored.
Furthermore, we note that there is an odd-even effect in the particle number
shown in Fig.~\ref{fig:S0_DE}, which is caused by the symmetry of the band
structure around $m_{\rm sym}=m_{\rm max}/2$ illustrated in
Fig.~\ref{fig:Band}, which leads to a kind
of shell structure in the single-particle energies.

Coulomb and Hubbard interaction strongly resemble each other at half filling
where the system is charge neutral and the physics is governed by the
short-ranged exchange interaction. Away from half filling, however, the
structure of the charged edges is dominantly determined by charge correlations
which are different for Coulomb or Hubbard interaction.

{\em Conclusions and outlook.}
We have presented a scheme to study the electronic
structure of midgap states at arbitrary filling, including the effects of
interactions. Examples are boundary states or 
states localized near vacancies, cracks, or wiggles, etc. close to
half-filling.  In our model all interactions are expressed in terms of midgap
states, and it is exact under the assumption of an inert, filled valence band.

We have applied our model to the localized states at zig-zag edges, where the
effects of interactions have been extensively studied, mostly using
short-ranged couplings like an onsite Hubbard term and mean field techniques.
We have not considered in detail effects due to the existence of two edges in
a graphene ribbon.  However, we note that the boundary
states of different edges are not coupled by Hubbard interaction. Treating the
kinetic coupling perturbatively one then finds that an antiferromagnetic
alignment of the two edges is favored, in agreement with previous
work~\cite{Son06,PCMH07,Brey07}. We expect the same behavior for Coulomb
interaction at half filling.

We found that the ground state close to half filling is spin-polarized for
both Hubbard and Coulomb interaction, which is in agreement with earlier
calculations. However, for a low electron (hole) occupation we
predict strong alternations of the total spin with the number of boundary electrons and
furthermore we find different behavior for Hubbard or Coulomb
interaction. The limits can be approximately described as a one
dimensional ferromagnet, near half filling, and a Wigner crystal
when the midgap states are almost empty.

In both limits, near half filling and an almost empty midgap band,
interactions lead to an effective one particle band which is dispersive, with
a bandwidth which has a well-defined limit when the length of the edge is much
larger than the lattice spacing. Hence, the boundary states acquire a finite
velocity, and can contribute to the transport properties of the system. We
have considered clean systems with no disorder. At half filling the low energy
states have a well-defined valley polarization and currents are valley
polarized at each edge. All possible backscattering processes therefore require
intervalley scattering, which is usually significantly smaller than
intravalley scattering~\cite{THGS08}.

For certain fillings (see $N_e=8$ in Fig.~\ref{fig:EMS}), the ground state is a doublet
with two possible values of the linear momentum. If the edge is shaped into a
ring, this result implies that, for these fillings, the ground state has a
persistent current of a mesoscopic size, as the current, $j \propto \partial E
/ \partial \phi$ where $\phi$ is an applied flux in a cylindrical geometry,
scales approximately as $R^{-1}$, where $R$ is the length of the edge.  As
discussed above, the decay of this current in samples with disorder is
controlled by intervalley scattering.

Our results also show that the spin of the lowest energy state depends
strongly on the total linear momentum of that state. This situation resembles
the dependence of the spin on total angular momentum in atoms and quantum
dots with circular or spherical symmetry.

{\em Acknowledgments}
We appreciate helpful discussions with L. Brey and S. Fratini. This work has been supported
by the EU Marie Curie RTN Programme No.  MRTN-CT-2003-504574, the EU Contract
12881 (NEST), and by MEC (Spain) through Grants No. MAT2002-0495-C02-01,
FIS2004-06490-C03-00, FIS2004-05120, FIS2005-05478-C02-01,FIS2007-65723, the Comunidad de
Madrid program CITECNOMIK, and the Juan de la Cierva Programme. T.S. is
supported by POCI 2010 via project PTDC/FIS/64404/2006 and by the ESF Science Program INSTANS.

\bibliography{qdotPaco}

\begin{thebibliography}{21}
\bibitem[{rmi()}]{rmin}
\bibinfo{note}{We set $(e^2/a)/U=0.45$ which for $e^2=\hbar v_F$ corresponds to
  $U=2t$.}
\end{thebibliography}
\end{document}